\documentclass[prb,preprint,groupedaddress]{revtex4-1}

\usepackage{graphicx, bm, amsfonts, amssymb, amsmath, appendix,figcaps,setspace}
\usepackage{color}
\newcommand{\T}{\mathrm{T}} 

\begin{document}

\title[Estimating averages using multiple time slices]
{Estimating equilibrium ensemble averages using multiple time slices from driven nonequilibrium processes:
theory and application to free energies, moments, and thermodynamic length in single-molecule pulling experiments}

\author{David D. L. Minh}
\affiliation{Biosciences Division, Argonne National Laboratory, 9700 S. Cass Av., Argonne, Illinois 60439, USA}
\email{daveminh@anl.gov}
\author{John D. Chodera}
\affiliation{California Institute for Quantitative Biosciences (QB3), University of California, Berkeley, Berkeley, California 94720, USA}
\email{jchodera@berkeley.edu}

\date{\today}

\begin{abstract}
Recently discovered identities in statistical mechanics have enabled the calculation of equilibrium ensemble averages from realizations of driven nonequilibrium processes, including single-molecule pulling experiments and analogous computer simulations.
Challenges in collecting large data sets motivate the pursuit of efficient statistical estimators that maximize use of available information.
Along these lines, Hummer and Szabo developed an estimator that combines data from multiple time slices along a driven nonequilibrium process to compute the potential of mean force.
Here, we generalize their approach, pooling information from multiple time slices to estimate arbitrary equilibrium expectations.
Our expression may be combined with estimators of path-ensemble averages, including existing optimal estimators that use data collected by unidirectional and bidirectional protocols.
We demonstrate the estimator by calculating free energies, moments of the polymer extension,  and the metric tensor for thermodynamic length in a model single-molecule pulling experiment.
Compared to estimators that only use individual time slices, our multiple time-slice estimators yield substantially smoother estimates and achieve lower variance for higher-order moments.
\begin{minipage}[c]{6.5 in}
\singlespace
\tiny
The submitted manuscript has been created by UChicago Argonne, LLC, Operator of Argonne National Laboratory (ÒArgonneÓ). Argonne, a U.S.ÊDepartment of Energy Office of Science laboratory, is operated under Contract No. DE-AC02-06CH11357. The U.S.ÊGovernment retains for itself, and others acting on its behalf, a paid-up nonexclusive, irrevocable worldwide license in said article to reproduce, prepare derivative works, distribute copies to the public, and perform publicly and display publicly, by or on behalf of the Government.
\end{minipage}
\end{abstract}

\maketitle

%%%%%%%%%%%%%%%%%%%%%%%%%%%%%%%%%%%%%%%%%%%%%%%%%%%%%%%%%%%%%%%%%%%%%
\section{Introduction}
%%%%%%%%%%%%%%%%%%%%%%%%%%%%%%%%%%%%%%%%%%%%%%%%%%%%%%%%%%%%%%%%%%%%%

When a system is driven out of equilibrium by a time-dependent external potential, the probability of finding it at a particular position in phase space generally differs from the equilibrium probability corresponding to the instantaneous thermodynamic state, a phenomenon known as \emph{lag}.  \cite{Pearlman1989}
With an appropriate reweighting of the nonequilibrium density, however, it is possible to recover an equilibrium distribution. \cite{Jarzynski1997b}
Ensemble averages can be computed by exploiting this fact;
the expected value of a phase-space dependent function, weighted by the dissipated work, is equal to an \emph{equilibrium} average of the same quantity (Eq.~\ref{eq:Crooks-expectation}).~\cite{Crooks2000, Neal2001}
This relationship, which is relevant to analyzing single-molecule pulling experiments and analogous computer simulations, has been applied to estimating various equilibrium properties of real and simulated systems. (See Ref.~\cite{MinhChodera2009} for a brief survey.)

Many asymptotically unbiased statistical estimators may be developed from this identity.
While these expressions will yield the same estimate in the limit of infinite sampling, their properties --- such as bias, variance, and smoothness --- will differ when applied to finite data.
Due to challenges in collecting large data sets, it is preferable to use statistically efficient estimators that have minimal bias and variance, and therefore make the best possible use of available information.

An implication of Eq.~\ref{eq:Crooks-expectation} is that equilibrium expectations may be estimated using data from \emph{any} time along a driven nonequilibrium process.
It is reasonable to surmise, however, that estimates of many properties will be improved by using data from \emph{all} recorded temporal observations, or \emph{time slices}.
For example, in a single-molecule pulling experiment, estimating the potential of mean force as a function of molecular extension typically involves creating a histogram of observed extensions.
The variance in this estimate can be enormous if the nonequilibrium density within a histogram bin is small.
Only by using a weighting scheme to combine data from multiple time slices were Hummer and Szabo able to produce a well-behaved estimator for the potential of mean force.~\cite{HummerSzabo2001}

The issue of stability (both numerical and statistical) in estimation from multiple time slices is surprisingly important.
Oberhofer and Dellago~\cite{Oberhofer2009} explored variations on Hummer and Szabo's weighting scheme,\cite{HummerSzabo2001}
deriving an alternative form which achieves lower variance in the limit of infinite sampling.
Unfortunately, correlations between time slices and the difficulty of accurately estimating covariance matrices from practically-sized samples, however, led to large fluctuations in the weights, and hence an unstable estimator.
In contrast, while Hummer and Szabo's approach \cite{HummerSzabo2001} is not asymptotically efficient, its balance between efficiency and robustness led to superior estimates in nearly all studied cases.~\cite{Oberhofer2009}  

Here, we present a previously unrecognized generalization of Hummer and Szabo's approach, applicable to the estimation of \emph{arbitrary} equilibrium expectations.
This generalization allows for multiple time-slice estimators to be constructed from any existing estimator for path-ensemble averages, such as optimized forms for unidirectional (the sample mean) or bidirectional data.\cite{MinhAdib2008, MinhChodera2009}
We then compare single and multiple time-slice forms in estimating free energies, moments of the polymer extension, and the thermodynamic length \cite{Crooks2007, Feng2009, Shenfeld2009} in a model single-molecule pulling experiment.

%%%%%%%%%%%%%%%%%%%%%%%%%%%%%%%%%%%%%%%%%%%%%%%%%%%%%%%%%%%%%%%%%%%%%
\section{Theory}
%%%%%%%%%%%%%%%%%%%%%%%%%%%%%%%%%%%%%%%%%%%%%%%%%%%%%%%%%%%%%%%%%%%%%

Consider a system evolving according to dynamics in which the stationary distribution of a configuration $x$ is given by
\begin{eqnarray}
\pi_\lambda(x) &=& Z_\lambda^{-1} \, q_\lambda(x) ,
\end{eqnarray}
where the partition function $Z_\lambda$ is,
\begin{eqnarray}
Z_\lambda &=& \int_\Gamma dx \, q_\lambda(x) ,
\end{eqnarray}
the unnormalized density $q_\lambda(x) = e^{-u_\lambda(x)}$ depends on the reduced potential~\cite{Shirts2008} $u_\lambda(x)$ 
(in which $\beta = (k_B T)^{-1}$ is absorbed into the potential) and satisfies $q_\lambda(x) \geq 0$ for all $x \in \Gamma$, and $\lambda$ is a \emph{vector} of one or more parameters that define the thermodynamic state.

Now suppose that the thermodynamic parameters $\lambda$ are varied in time according to the protocol $\Lambda \equiv \{\lambda_0, ..., \lambda_T\}$ over $t \in \{0, \ldots, T\}$.
At each time slice $t$, the system evolves according to dynamics which preserve the distribution $\pi_{\lambda_t}$.
For notational convenience, we henceforth write $\pi_t(x)$ instead of $\pi_{\lambda_t}(x)$, $u_t(x)$ instead of $u_{\lambda_t}(x)$, and $Z_t$ instead of $Z_{\lambda_t}$.

Let $E_{0 \rightarrow t}[\mathcal A]$ denote the nonequilibrium expectation of a path functional $\mathcal A[X]$ over all possible realizations $X \equiv \{x_0, ..., x_T\}$ of a process starting with $x_0$ drawn from the equilibrium distribution $\pi_0(x)$.
We also define the equilibrium expectation of a function $A(x)$ with respect to the equilibrium density $\pi_t(x)$ as $E_t[A] \equiv \int_\Gamma dx \, A(x) \pi_t(x)$.
With these definitions, the following identity holds for all $t \in \{0, \ldots, T\}$:~\cite{Crooks2000, Neal2001}
\begin{eqnarray}
\label{eq:Crooks-expectation}
E_t[A] &=& E_{0 \rightarrow t} \left[A(x_t) \, e^{-w_{0 \rightarrow t}} \right] \frac{Z_0}{Z_t} ,
\end{eqnarray}
where $w_{0 \rightarrow t}[X]$ denotes the appropriate work function for the switching process, \cite{Crooks2000, Neal2001, Ritort2009}
\begin{eqnarray}
w_{0 \rightarrow t}[X] &\equiv& \sum_{t=1}^T \left[ u_t(x_t) - u_{t-1}(x_t) \right] .
\end{eqnarray}

Depending on how paths are sampled, the nonequilibrium expectation $E_{0 \rightarrow t}[\mathcal A]$ may be estimated via a number of methods. 
In this paper, we use the notation $\mathcal E_{0 \rightarrow t}[\mathcal A]$ to denote an \emph{estimator} of the nonequilibrium path expectation $E_{0 \rightarrow t}[\mathcal A]$ that makes use of finite data, $\mathcal E_t[A]$ an estimator for the equilibrium expectation $E_t[A]$, and $\hat{Z}_\lambda$ to denote an estimator of $Z_\lambda$ up to an arbitrary multiplicative constant that is identical for all $\lambda$.

Suppose $N_f$ paths are sampled from a single protocol.  With these sampled paths denoted as $X_{fn}, n = 1, ..., N_f$, the most appropriate estimator is the sample mean of the path functional over the sampled paths,	
\begin{eqnarray}
\mathcal{E}_{0 \rightarrow t}[\mathcal{A}] = \frac{1}{N_f} \sum_{n=1}^{N_f} \mathcal{A}[X_{fn}] . 
\label{eq:unidirectional-path-estimator}
\end{eqnarray}
When paths are also sampled according to the \emph{reverse} process, $\tilde{\Lambda} \equiv \{\tilde{\lambda}_0, ..., \tilde{\lambda}_T\} = \{\lambda_T, ...,\lambda_0\}$, and the dynamics at fixed $\lambda$ satisfy detailed balance, the estimator, \cite{MinhAdib2008}
\begin{eqnarray}
\mathcal{E}_{0 \rightarrow t}[\mathcal A] &=& \sum_{n=1}^{N_f} \frac{ \mathcal A[X_{fn}] }{ N_f + N_r \, (\hat{Z}_0/\hat{Z}_T) \, e^{-w_{0 \rightarrow t}[X_{fn}] } } + 
\sum_{m=1}^{N_r} \frac{ \mathcal A[X_{rm}] }{ N_f + N_r \, (\hat{Z}_0/\hat{Z}_T) \, e^{-w_{0 \rightarrow t}[X_{rm}] } }, \label{eq:bidirectional-path-estimator}
\end{eqnarray}
has been shown to be asymptotically efficient~\cite{MinhChodera2009} when the ratio $\hat{Z}_0 / \hat{Z}_T$ is estimated by choosing $\mathcal{A}[X] \equiv 1$, which yields the well-known Bennett acceptance ratio.~\cite{Bennett1976, Shirts2003}
Here, $X_{rm}$ denotes the \emph{time reversal}~\cite{Crooks1998, Jarzynski2006, MinhAdib2008} of a path generated using the protocol $\tilde{\Lambda}$, indexed according to $m = 1, ..., N_r$. 
Optimal estimators relevant to trajectories sampled from multiple path-ensembles have also been described.~\cite{MinhChodera2009}

%%%
\subsection{Single time-slice estimators}
%%%

Note that for any time slice $t$, we can obtain an expectation with respect to an arbitrary $\pi_*(x)$ using the importance sampling identity,
\begin{eqnarray}
\label{eq:isamp}
E_*[A] &=& E_{t}\left[ A(x) \, \frac{\pi_*(x)}{\pi_{t}(x)} \right].
\end{eqnarray}
Substituting Eq.~\ref{eq:Crooks-expectation}, this may be expressed in terms of an average over nonequilibrium paths as,
\begin{eqnarray}
\label{eq:single-time-expectation}
E_*[A] &=& E_{0 \rightarrow t}\left[ A(x_t) e^{-w_{0 \rightarrow t}} \, \frac{q_*(x_t)}{q_t(x_t)} \right] \frac{Z_0}{Z_*}.
\end{eqnarray}
Replacing the expectations with their corresponding estimators, we obtain,
\begin{eqnarray}
\label{eq:single-time-estimator}
\mathcal E_*[A] = 
\mathcal E_{0 \rightarrow t}\left[ A(x_t) \, e^{-w_{0 \rightarrow t}} \, \frac{q_*(x_t)}{q_t(x_t)} \right] \frac{\hat{Z}_0}{\hat{Z}_*}.
\end{eqnarray}
Use of this expression to estimate arbitrary expectations $E_*[A]$ requires an estimate of the unknown ratio $\hat{Z}_0 / \hat{Z}_*$.
While in theory there exist several means of estimating this ratio, 
one important criterion for choosing an estimator is the self-consistency of Eq.~\ref{eq:single-time-estimator};
it is necessary for estimates of constant functions $A(x) = C$ to yield the same value, $C$.
As not all estimators for $\hat{Z}_0 / \hat{Z}_*$ will properly balance the weighing factors in Eq.~\ref{eq:single-time-estimator} and satisfy this criterion, there is a constraint on possible estimates of the ratio.
Fortunately, the choice $A(x) = 1$ in Eq.~\ref{eq:single-time-expectation} leads to the convenient estimator,
\begin{eqnarray}
\label{eq:estimator-for-normalization-constants}
\frac{\hat{Z}_*}{\hat{Z}_0} &=& \mathcal E_{0 \rightarrow t}\left[ e^{-w_{0 \rightarrow t}} \, \frac{q_*(x_t)}{q_t(x_t)} \right].
\end{eqnarray}
When $\lambda_* = \lambda_t$, this choice of estimator for $\hat{Z}_0 / \hat{Z}_*$ is equivalent to the single time-slice estimator based on 
Jarzynski's equality. \cite{Jarzynski1997a, Jarzynski1997b, MinhChodera2009}

%%%
\subsection{Multiple time-slice estimators}
%%%

Eq.~\ref{eq:single-time-estimator} only uses configuration data from a single time slice, $t$.
For some observables $A(x)$, a number of time slices may contain information relevant to the estimation of $E_*[A]$.
To combine data from multiple time slices in a stable manner, we consider the multiple importance sampling (MIS) strategy of Guibas and Veach.~\cite{Veach1995, Veach1997}
As in earlier work focusing on potentials of mean force,\cite{Oberhofer2009} 
this strategy is implemented by introducing a weight function $\alpha_t(x)$, subject to the conditions $\alpha_t(x) \ge 0$ for all $x$ and $t$, and the constraint $\sum_{t=0}^T \alpha_t(x) = 1$ for all $x$.
By introducing these weights into Eq.~\ref{eq:isamp} through a factor of unity and applying Eq.~\ref{eq:Crooks-expectation}, we obtain the identity,
\begin{eqnarray}
E_*[A] &=& E_*\left[ \left( \sum_{t=0}^T \alpha_t(x) \right) A(x) \right] \nonumber \\
&=& \sum_{t=0}^T E_*\left[ \alpha_t(x) \, A(x) \right] \nonumber \\
&=& \frac{Z_0}{Z_*} \sum_{t=0}^T E_{0 \rightarrow t}\left[ \alpha_t(x_t)\, A(x_t) \, e^{-w_{0 \rightarrow t}} \, \frac{q_*(x_t)}{q_t(x_t)} \right].
\end{eqnarray}
By replacing the above expectations with estimators, we obtain the general form of the MIS estimator for equilibrium expectations that uses multiple time slices from driven nonequilibrium processes,
\begin{eqnarray}
\label{eq:general-MIS-estimator}
\mathcal E_*[A] &=& \frac{\hat{Z}_0}{\hat{Z}_*} \sum_{t=0}^T \mathcal E_{0 \rightarrow t}\left[ \alpha_t(x_t)\, A(x_t) \, e^{-w_{0 \rightarrow t}} \, \frac{q_*(x_t)}{q_t(x_t)} \right].
\end{eqnarray}
Eq.~\ref{eq:general-MIS-estimator} can be seen as a generalized form of Eq.~8 from Oberhofer and Dellago,~\cite{Oberhofer2009} applicable not only to potentials of mean force, but to \emph{arbitrary} expectations.  
(Applications of this general form to quantities obtainable from single-molecule pulling experiments, including potentials of mean force, are described in section \ref{sec:single-molecule}.)

Every weighting function $\alpha_t(x)$ that satisfies the above conditions results in an asymptotically consistent estimator that produces the true expectation in the limit of infinite data, but will have different properties for finite sample sizes.
The choice $\alpha_t(x) \equiv \delta_{tt_*}$, where $\delta_{ij}$ denotes the Kronecker delta and $t_*$ a designated time slice, recapitulates the single time-slice estimator of Eq.~\ref{eq:single-time-estimator}.
Another possibility is to weight all time slices equally by choosing $\alpha_t(x) \equiv (T+1)^{-1}$,
\begin{eqnarray}
\label{eq:MIS-estimator-uniform}
\mathcal E_*[A] &=& \frac{\hat{Z}_0}{\hat{Z}_*} \left(\frac{1}{T+1}\right) \sum_{t=0}^T \mathcal E_{0 \rightarrow t}\left[ A(x_t) \, e^{-w_{0 \rightarrow t}} \, \frac{q_*(x_t)}{q_t(x_t)} \right].
\end{eqnarray}
As in Eq.~\ref{eq:estimator-for-normalization-constants}, the choice $A(x) = 1$ leads to an estimator for the required ratio $\hat{Z}_* / \hat{Z}_0$,
\begin{eqnarray}
\label{eq:MIS-estimator-for-normalization-constants-uniform}
\frac{\hat{Z}_*}{\hat{Z}_0} &=& \left(\frac{1}{T+1}\right) \sum_{t=0}^T \mathcal E_{0 \rightarrow t}\left[ e^{-w_{0 \rightarrow t}} \, \frac{q_*(x_t)}{q_t(x_t)} \right].
\end{eqnarray}
Unfortunately, this choice is expected to perform poorly; time slices that carry little information about $E_*[A]$, because their instantaneous nonequilibrium densities $\rho_{0 \rightarrow t'}(x) \equiv E_{0 \rightarrow t'}[\delta(x - x_{t'})]$ differ greatly from $\pi_*(x)$, are treated equally to those that carry the most.
(This na\"{\i}ve weighting scheme has previously been used to estimate a potential of mean force. \cite{Paramore2007})

A more stable choice that makes better use of all time slices weights the contribution from configuration $x$ according to its \emph{equilibrium} probability:
\begin{eqnarray}
\alpha_t(x) &=& \frac{\pi_t(x)}{\sum\limits_{t'=0}^T \pi_{t'}(x)}.
\end{eqnarray}
This choice corresponds to the balance heuristic~\cite{Veach1997} of MIS, and leads to the estimator,
\begin{eqnarray}
\label{eq:MIS-estimator-balance}
\mathcal E_*[A] &=& \frac{\hat{Z}_0}{\hat{Z}_*} \sum_{t=0}^T \mathcal E_{0 \rightarrow t}\left[ \frac{ \hat{Z}_t^{-1} \, q_*(x_t) }{\sum\limits_{t'=0}^T \hat{Z}_{t'}^{-1} q_{t'}(x_t)} \, A(x_t) \, e^{-w_{0 \rightarrow t}} \right].
\end{eqnarray}
Again, we obtain an estimator for the ratio $\hat{Z}_* / \hat{Z}_0$ by choosing $A(x)=1$,
\begin{eqnarray}
\frac{\hat{Z}_*}{\hat{Z}_0} &=& \sum_{t=0}^T \mathcal E_{0 \rightarrow t}\left[ \frac{ \hat{Z}_t^{-1} \, q_*(x_t) }{\sum\limits_{t'=0}^T \hat{Z}_{t'}^{-1} q_{t'}(x_t)} \, e^{-w_{0 \rightarrow t}} \right].
\label{eq:MIS-estimator-for-normalization-constants-balance}
\end{eqnarray}
As a word of caution, we note that use of estimators for $\hat{Z}_* / \hat{Z}_0$ other than Eq.~\ref{eq:MIS-estimator-for-normalization-constants-balance} (such as Eq.~\ref{eq:MIS-estimator-for-normalization-constants-uniform}) will lead to a violation of the imposed constraint that the estimated expectation of a constant function is a constant.
In other words, Eq.~\ref{eq:MIS-estimator-for-normalization-constants-balance} must be used with Eq.~\ref{eq:MIS-estimator-balance}, while Eq.~\ref{eq:MIS-estimator-for-normalization-constants-uniform} must be used with Eq.~\ref{eq:MIS-estimator-uniform}.

\subsection{Thermodynamic length}
%%%

One possible application of these estimators is the calculation of thermodynamic length, 
a natural measure of distance on the manifold of equilibrium thermodynamic states 
(see Ref. ~\cite{Crooks2007} for an excellent overview of the concept).
Thermodynamic length is related to the heat dissipated during endoreversible processes,~\cite{SalamonBerry1983} 
%(in which the system is in equilibrium except for heat exchange with the surroundings)
and may prove useful in the optimization of fractional distillation,~\cite{SalamonNulton1998}
the design of molecular motors,~\cite{Feng2009}
and the selection of efficient data collection protocols.~\cite{Shenfeld2009}

The thermodynamic length of a continuous protocol
$\Lambda \equiv \lambda(t)$ in parameter space is defined as a path integral of the thermodynamic metric tensor $\mathcal{I}_{\lambda}$,
\begin{eqnarray}
\mathcal{L} &\equiv& \int_0^\tau dt \, \left( \dot{\lambda} \cdot \mathcal{I}_{\lambda} \cdot \dot{\lambda} \right)^{1/2},
\end{eqnarray}
where $\dot{\lambda} \cdot \mathcal{I}_{\lambda} \cdot \dot{\lambda}$ denotes a vector-tensor-vector inner product (in the case of multidimensional thermodynamic parameters $\lambda$) and $\dot{\lambda}$ denotes the time derivative of $\lambda(t)$. 
The metric tensor $\mathcal{I}_\lambda$ is the Fisher information matrix \cite{Fisher1925} on the manifold of equilibrium thermodynamic states,
\begin{eqnarray}
\mathcal{I}_{\lambda} &=& E_\lambda\left[\left(\nabla_\lambda \ln \pi_\lambda\right) \, \left(\nabla_\lambda \ln \pi_\lambda\right)^\T \right] \nonumber ,
\end{eqnarray}
where $x y^\T$ denotes the outer product between vectors $x$ and $y$.
In most situations, the thermodynamic length cannot be computed directly; instead, it can be approximated by numerical quadrature using estimates of  $\mathcal{I}_\lambda$ computed at discrete points along $\lambda(t)$.

An alternative strategy is to compute the discrete-time analogue of the Fisher length, the Jensen-Shannon length, \cite{Feng2009}
\begin{eqnarray}
\mathcal L_{JS} \equiv \sqrt{8} \sum_{t=0}^{T-1} \sqrt{\mathcal D_{JS}(\pi_t(x),\pi_{t+1}(x))},
\label{eq:JSlength}
\end{eqnarray}
which contains the Jensen-Shannon divergence, \cite{Lin1991}
\begin{eqnarray}
\mathcal D_{JS}(\pi_j,\pi_k) & \equiv &
\frac{1}{2} E_{j} \left[ \ln \frac{\pi_j}{\frac{1}{2}(\pi_j(x)+\pi_k(x))} \right] + 
\frac{1}{2} E_{k} \left[ \ln \frac{\pi_k}{\frac{1}{2}(\pi_j(x)+\pi_k(x))} \right] 
\end{eqnarray}
The Jensen-Shannon length satisfies $\mathcal L_{JS} \leq \mathcal L$, approaching equality as the step size decreases. \cite{Crooks2007}
Estimators for the thermodynamic length based on the Jensen-Shannon divergence have been previously derived,~\cite{Crooks2007, Feng2009} but not tested on any data, simulated or experimental.
In this paper we compare estimators on a model of a single-molecule pulling experiment.

%%%
\subsection{Application to single-molecule pulling experiments\label{sec:single-molecule}}
%%%

Estimators of equilibrium ensemble averages from driven nonequilibrium processes are particularly relevant to single-molecule pulling experiments.
Indeed, single-molecule force spectroscopy has been used to experimentally verify theorems relating nonequilibrium processes with equilibrium properties. \cite{Liphardt2002, Collin2005} 
These theorems have also been applied to computing RNA folding free energies as a function of a control parameter. \cite{Junier2009}
Here, we specifically consider an experiment in which two polystyrene beads are attached to a polymer, such as a nucleic acid or protein.
One bead is held at the origin, affixed to a micropipette, and the other is held in an optical trap centered about position $\bar{z}(t)$ along the $z$-axis.

The total reduced potential of this system at inverse temperature $\beta$ is described by 
\begin{eqnarray}
u_t(x) &=& u_b(x) + v_t(z(x)) \nonumber
\end{eqnarray}
where $u_b(x)$ is the bare reduced potential, and 
\begin{eqnarray}
v_t(z(x)) &=& \frac{\beta k_s}{2} (z(x) - \bar{z}(t))^2
\end{eqnarray}
is the harmonic biasing reduced potential with spring constant $k_s$ associated with the optical trap (e.g.~Fig.~\ref{fig:potential}).

In the absence of the external harmonic biasing potential $v_t(z)$, 
the potential of mean force (PMF) along the $z$-axis is given by
\begin{eqnarray}
g_b(z) &\equiv& - \ln E_b[\delta(z(x) - z)] + \delta g,
\end{eqnarray}
where $\delta g$ is an arbitrary constant.
This PMF may be estimated using Eq.~\ref{eq:MIS-estimator-balance}, leading to,
\begin{eqnarray}
\lefteqn{e^{-g_b(z) + \delta g} =} \nonumber \\
&=& \frac{\hat{Z}_0}{\hat{Z}_b} \sum_{t=0}^T \mathcal E_{0 \rightarrow t}\left[ \frac{ \hat{Z}_t^{-1} \, e^{-u_b(x_t)} }{\sum\limits_{t'=0}^T \hat{Z}_{t'}^{-1} e^{-u_t(x_t) }} \, \delta(z(x_t) - z) \, e^{-w_{0 \rightarrow t}} \right] \nonumber \\
&=& \frac{\hat{Z}_0}{\hat{Z}_b} \frac{\sum\limits_{t=0}^T \left( \frac{\hat{Z}_0}{\hat{Z}_t} \right) \, \mathcal E_{0 \rightarrow t}\left[ \delta(z(x_t) - z) \, e^{-w_{0 \rightarrow t}} \right]}{\sum\limits_{t'=0}^T \left( \frac{\hat{Z}_0}{\hat{Z}_{t'}} \right) e^{-v_{t'}(z)}}.
\label{eq:MIS-pmf-estimator}
\end{eqnarray}
Defining $e^{\delta g} \equiv \hat{Z}_0 / \hat{Z}_b$, we obtain precisely Eq.~8 from  Hummer and Szabo~\cite{HummerSzabo2001} and Eq.~9 from Minh and Adib.~\cite{MinhAdib2008}

As properties of Eq.~\ref{eq:MIS-pmf-estimator} have been examined in detail elsewhere,~\cite{HummerSzabo2001, MinhAdib2008}
here we concentrate on the comparison of estimators for other equilibrium averages:
\begin{enumerate}
\item Free energies of the entire system, including the harmonic potential, $F_0^t \equiv -\ln \left( Z_t / Z_0\right)$;
\item Moments of $z$ about the mean, $E_t[(z - E_t[z])^n]$, for $n = 1,...,6$;
\item The metric tensor $\mathcal I_t$ associated with thermodynamic length;
\end{enumerate}
for all $t \in \{0, \ldots, T\}$.

In the single-molecule pulling experiment considered here, 
the trap position is the only thermodynamic parameter which is varied, 
and hence the Fisher information matrix contains a single element,
\begin{eqnarray}
\mathcal{I}_t = (\beta k_s)^2 E_t[(z - E_t[z])^2],
\end{eqnarray}
which is proportional to the second central moment of the polymer extension, 
a quantity observable in single-molecule pulling experiments as well as computer simulations.
Hence, all of these quantities may be estimated in both laboratory experiments and computer simulations.

%%%%%%%%%%%%%%%%%%%%%%%%%%%%%%%%%%%%%%%%%%%%%%%%%%%%%%%%%%%%%%%%%%%%%
\section{Illustrative Example}
%%%%%%%%%%%%%%%%%%%%%%%%%%%%%%%%%%%%%%%%%%%%%%%%%%%%%%%%%%%%%%%%%%%%%

\begin{figure}[tb]
\includegraphics{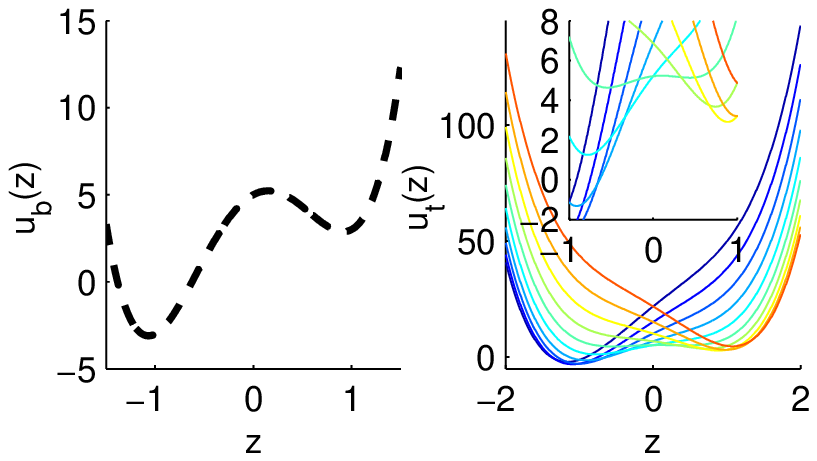}
\caption{{\bf Model potential for a single-molecule pulling experiment.}
Left: Bare reduced potential $u_b(z)$.
Right: Total reduced potential, including the external harmonic biasing potential, $u_t(z)$ at ten different times spanning from $\bar{z}(t) = -1.5$ (blue) to $+1.5$ (red).  
Inset: same total reduced potential, zoomed in.
Potential energies are shown in units of $k_BT$.
\label{fig:potential}} 
\end{figure}

As a model of a single-molecule pulling experiment, consider a one-dimensional double-well system with a bare (unbiased) reduced potential $u_b(z) = (5z^3 - 10z + 3)z$ (Fig.~\ref{fig:potential}).  We perform overdamped Langevin (Brownian) dynamics simulations on this system as previously described.~\cite{MinhAdib2008, MinhChodera2009}
The total reduced potential $u_t(z) = u_b(z) + v_t(z)$ includes a harmonic biasing potential $v_t(z) = \frac{\beta k_s}{2} (z - \bar{z}(t))^2$ with reduced spring constant $(\beta k_s) = 15$.
The position is propagated using $z_{t+1} = z_t - D(\partial/\partial x)u_t(x_t) \Delta t + (2D\Delta t)^{1/2} R_t$, where the diffusion coefficient is $D = 1$, the time step is $\Delta t = 0.001$, 
and $R_t \sim N(0,1)$ is a sequence of independent, identically distributed random numbers drawn from the standard normal distribution.
After equilibration at the initial $\bar{z}$, 250 pulling trajectories were performed in both the forward ($\bar{z}(t) = -1.5 + 0.004 t$) and reverse ($\bar{z}(t) = +1.5 - 0.004 t$) directions for 750 steps.
Equilibrium ensemble averages were estimated in three ways: using only the forward trajectories (the \emph{forward} experiment), only the reverse trajectories (the \emph{reverse} experiment), or half of the trajectories from the forward and reverse ensembles (the \emph{bidirectional} experiment).
To assess the variance and bias of the estimates, independent experiments were replicated 2500 times and statistics accumulated to compare the bias and variance of the different estimators.
Reference values of the free energies $F_0^t$, moments about the mean, the metric tensor $\mathcal{I}_t$, the Jensen-Shannon divergence $\mathcal D_{JS}(\pi_t,\pi_{t+1})$, and the thermodynamic length $\mathcal L$, were numerically computed by adaptive Gauss-Kronrod quadrature using the {\tt quadgk} method provided in MATLAB 7.10.0.499 (R2010a).

\begin{figure}[tb]
\includegraphics{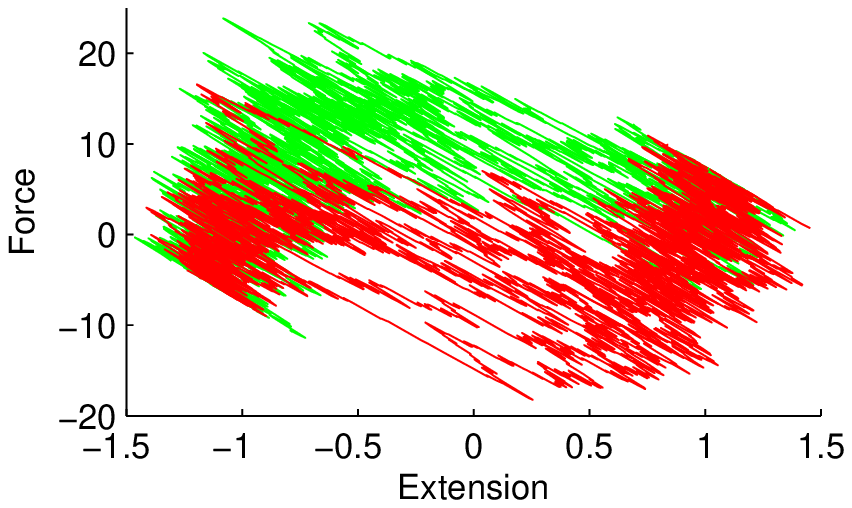}
\caption{{\bf Force-Extension Curves.}
Ten representative force-extension curves from forward (green) and reverse (red) pulling simulations.
\label{fig:force_extension}} 
\end{figure}

As can be seen in the force-extension curves (Fig.~\ref{fig:force_extension}), the chosen pulling speed is sufficient to introduce significant hysteresis into the system.
While approximately the same range of forces and extensions are sampled near the beginning and ends of the forward and reverse pulling simulations, the barrier-crossing forces in the forward direction are generally higher than in the reverse.  
If the pulling speed is increased, the extent of hysteresis increases.  
Conversely, if it is decreased, forward and reverse trajectories (after appropriate time reversal) are less distinguishable (data not shown).  
This speed was chosen to be slow enough for estimates of equilibrium quantities to converge, but fast enough so that performance differences between unidirectional and bidirectional estimators are evident.

\begin{figure}[tb]
\includegraphics{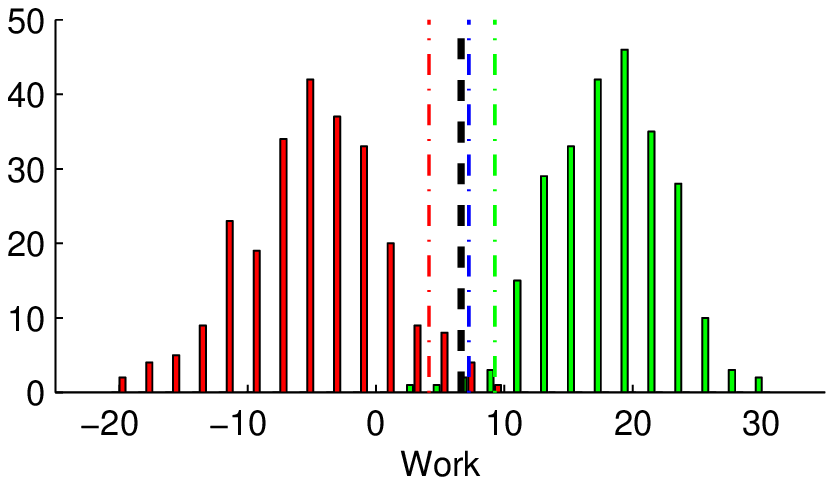}
\caption{{\bf Work histograms.}
Representative histograms of work performed in forward trajectories (green) and negative work in the reverse trajectories (red).   The free energy difference between states with the harmonic trap at $\bar{z}(t) = -1.5$ and $\bar{z}(t) = +1.5$, computed by numerical quadrature, is shown as a thick dashed black line.  Estimates of the free energy difference from 250 forward (green) or 250 reverse (red) using Jarzynski's equality \cite{Jarzynski1997a, Jarzynski1997b, MinhChodera2009} or 125 pulling simulations in each direction using the Bennett Acceptance Ratio (blue) \cite{Bennett1976, Crooks2000} are shown as lines alternating between dashed and dotted symbols.
\label{fig:workdist}} 
\end{figure}

The effect of hysteresis is also evident in the work histograms (Fig.~\ref{fig:workdist}), which are fairly broad.
Indeed, the extent of dissipation makes it difficult to estimate the free energy difference using Jarzynski's equality; with this example, the unidirectional estimates in both the forward and reverse directions are overestimated.
According to the Crooks fluctuation theorem, \cite{Crooks1998} the work distribution in the forward direction and the negative work in the reverse direction should cross at the free energy difference.
While this crossing point is difficult to pinpoint by examining histograms (as was done in several recent analyses of single-molecule pulling experiments), a free energy estimate based on the Bennett Acceptance Ratio \cite{Bennett1976,Crooks2000} is fairly accurate.

\begin{figure}[tb]
\includegraphics{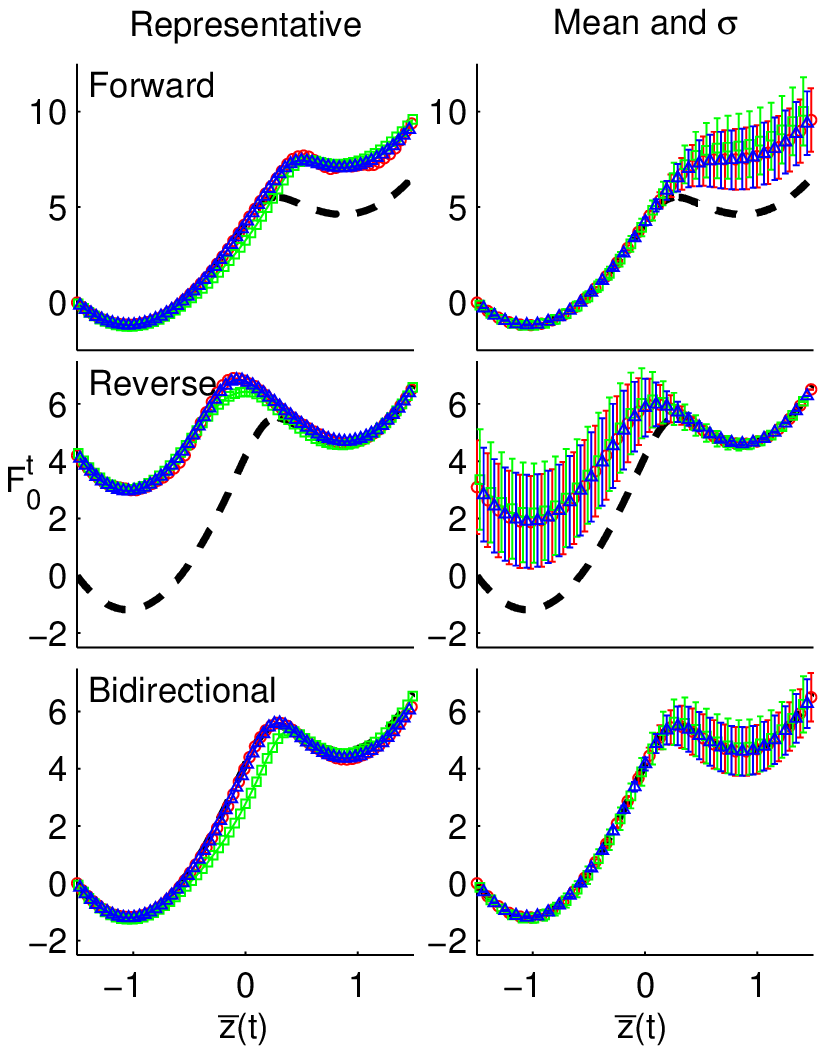}
\caption{{\bf Estimates of free energy differences.}
Representative estimates (left column) and the mean and standard deviation of 2500 estimates (error bars, right column) of $F_0^t$ shown as a function of $\bar{z}(t)$ from $-1.5$ to $+1.5$.
Estimates were computed with the single time-slice estimator, Eq.~\ref{eq:estimator-for-normalization-constants} (red circles),
the MIS estimator with uniform weighting, Eq.~\ref{eq:MIS-estimator-for-normalization-constants-uniform} (green squares),
and the MIS estimator with the balance heuristic, Eq.~\ref{eq:MIS-estimator-for-normalization-constants-balance} (blue triangles), 
utilizing only 250 forward (top), only 250 reverse (middle), or 125 pulling simulations in each direction (bottom).
For improved clarity, not all points are shown.
The $F_0^t$ computed by numerical quadrature is shown as a thick dashed black line.
All free energies are shown in units of $k_BT$.
\label{fig:F0t}} 
\end{figure}

We have considered the free energy estimates not only at the end points, but as a function of trap position (Fig.~\ref{fig:F0t}).
In these calculations, the performance (bias and variance) of the MIS estimator is not substantially different from the single time-slice estimator (Fig.~\ref{fig:F0t}).
Unidirectional estimates, as previously noted, \cite{MinhAdib2008, MinhChodera2009}
are increasingly biased and have larger variance as the system is driven further away from equilibrium.
Using multiple time slices does not alleviate this situation.
Indeed, the MIS estimator with uniform weight, Eq.~\ref{eq:MIS-estimator-for-normalization-constants-uniform}, has a slightly greater bias and variance than the single time-slice estimator, Eq.~\ref{eq:estimator-for-normalization-constants}.
Using the balance heuristic, Eq.~\ref{eq:MIS-estimator-for-normalization-constants-balance}, leads to an estimator with very similar performance.
The similarity of the estimates is likely due to the high degree of correlation in the work values from sequential time slices.
This does not preclude the possibility, however, that the multiple time-slice estimator will perform better than the single time-slice estimator in other systems.

\begin{figure}[tb]
\includegraphics{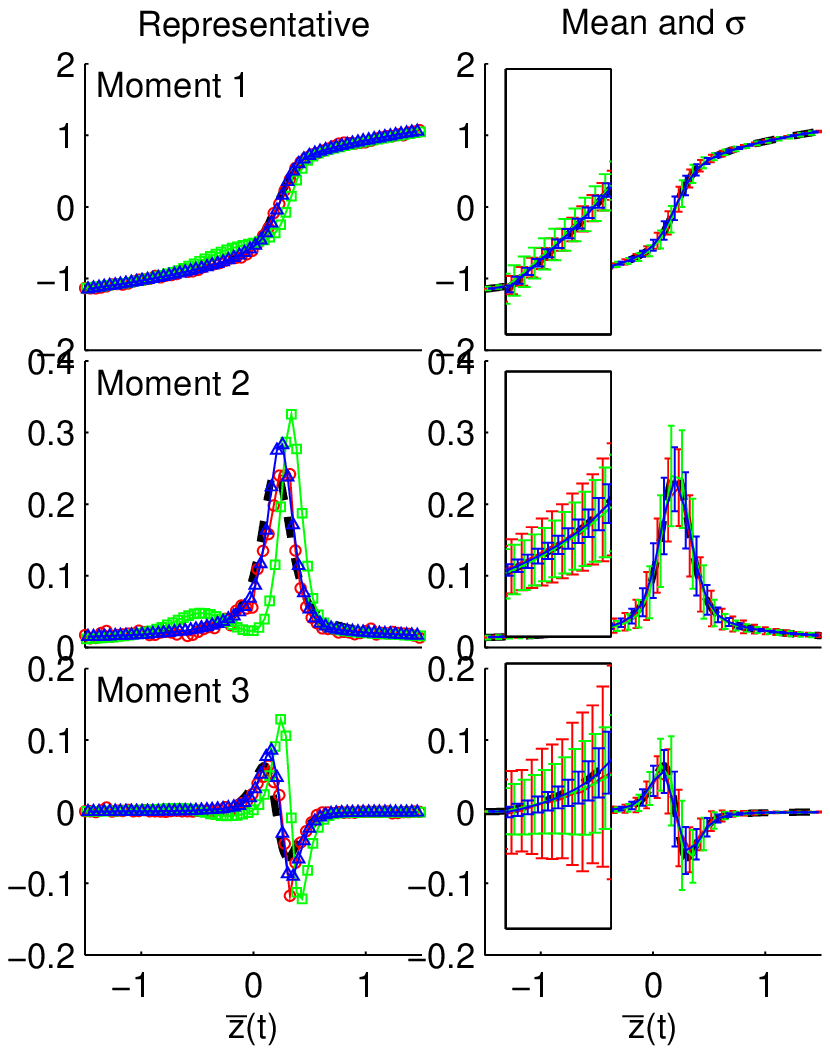}
\caption{{\bf Estimates of moments of $z$ about the mean.}
Representative estimates (left column) and the mean and standard deviation of 2500 estimates (error bars, right column) of the first (top), second (middle), and third (bottom) central moments, $E_t[(z - E_t[z])^n]$, 
shown as a function of $\bar{z}(t)$ from $-1.5$ to $+1.5$.
Estimates were computed with the single time-slice estimator, Eq.~\ref{eq:estimator-for-normalization-constants} (red circles or error bars), 
the MIS estimator with uniform weighting, Eq.~\ref{eq:MIS-estimator-for-normalization-constants-uniform} (green squares or error bars), 
and the MIS estimator with the balance heuristic, Eq.~\ref{eq:MIS-estimator-for-normalization-constants-balance} (blue triangles or error bars), 
utilizing 125 pulling simulations in each direction.
The inset shows a closer view of the left tail of the mean and standard deviation, at a range between $\bar{z}(t) = -1$ and $-0.5$.
For improved clarity, not all points are shown.
Moments computed by numerical quadrature are shown as thick dashed black lines.
\label{fig:moments123}} 
\end{figure}

\begin{figure}[tb]
\includegraphics{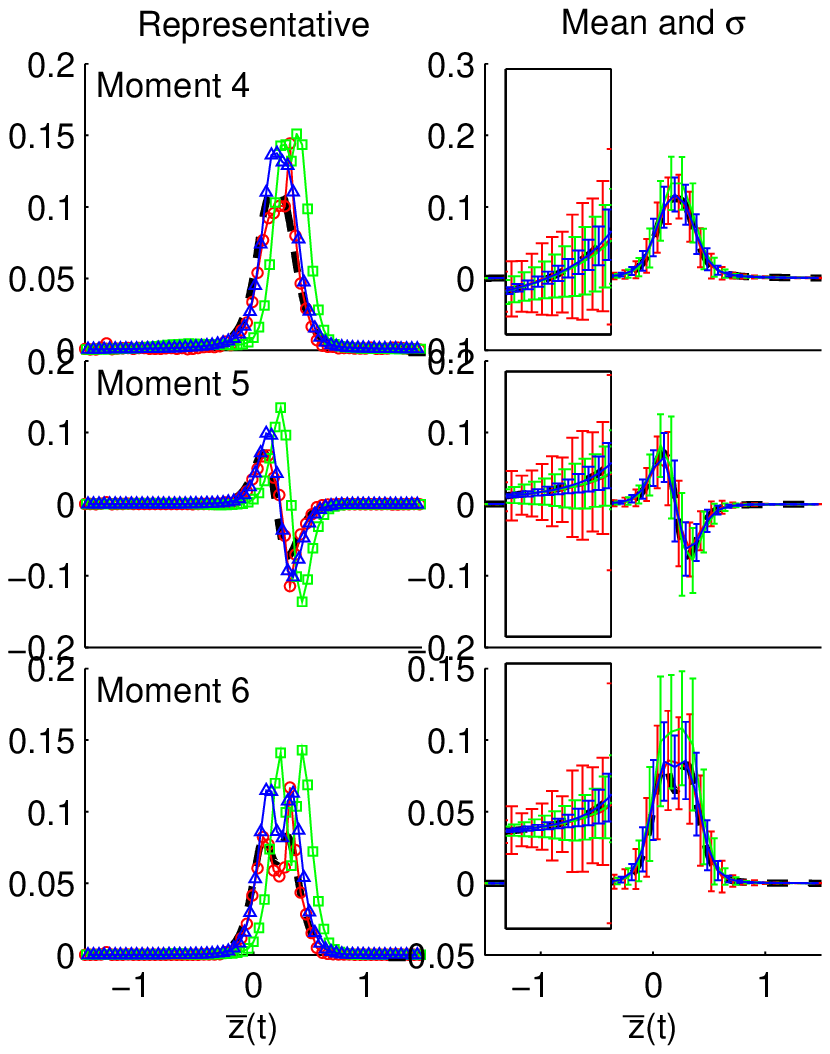}
\caption{{\bf Estimates of moments of $z$ about the mean.}
Representative estimates (left column) and the mean and standard deviation of 2500 estimates (error bars, right column) of the fourth (top), fifth (middle), and sixth (bottom) moments about the mean.  Otherwise, the caption in Fig.~\ref{fig:moments123} applies here.
\label{fig:moments456}} 
\end{figure}

On the other hand, results from different methods of estimating moments about the mean are more distinct (Figs.~\ref{fig:moments123} and \ref{fig:moments456}), and the disparities are larger for higher-order moments.
(Because of the large bias in unidirectional estimates, we have only shown results from bidirectional estimates.)
With the first moment, the bias and variance properties of the methods are quite similar, except for the MIS estimator with uniform weighting, which performs slightly worse (top right of Fig.~\ref{fig:moments123}).
As the average of a large number (e.g.~2500) of estimates is likely closer to the true value than any individual estimate, it is also informative to examine single estimates.
In this case, the single time-slice estimator has fluctuations which are somewhat misleading, since the actual first-order moment varies smoothly with $\bar{z}$.
The MIS estimator with the balance heuristic, on the other hand, has a smoothness which more accurately reflects the true value.

For second- and higher-order moments, the benefits of the MIS estimator are more pronounced.
In addition to the previously observed trends in smoothness (left column of Figs.~\ref{fig:moments123} and \ref{fig:moments456}),
the variance of various estimators is significantly different (right column of Figs.~\ref{fig:moments123} and \ref{fig:moments456}).
Compared to the single time-slice estimator, the MIS estimator with the balance heuristic has substantially reduced variance, especially at the extreme values of $\bar{z}$.
Results with uniform weighting are mixed, as performance is improved at the extreme values but is worse in the middle of the distribution.
These results not only demonstrate the value of pooling information from multiple time slices, but in using a high-quality weighting procedure.

%\begin{widetext}
Now consider thermodynamic length.
Since the metric tensor for the thermodynamic length $\mathcal I_t$ is directly proportional to the variance, the above comparison of single and multiple time-slice estimators holds;
the MIS estimator with balance heuristic produces estimates with much smaller statistical error than the single time-slice estimator.
An alternate strategy for measuring thermodynamic length, based on the 
Jensen-Shannon divergence, also merits comparison.
Feng and Crooks noted that the Jensen-Shannon divergence between equilibrium probability distributions along the protocol of a driven nonequilibrium process is related to the sum of two path-ensemble averages, \cite{Feng2009}
\begin{eqnarray}
\mathcal D_{JS}(\pi_t,\pi_{t+1}) & = &
\frac{1}{2} E_{0 \rightarrow T} \left[ \frac{Z_0}{Z_t} e^{-w_{0 \rightarrow t}} \ln \frac{2}{1 + \frac{Z_t}{Z_{t+1}} e^{-w_{t \rightarrow t+1}}} \right] \nonumber \\
&+& \frac{1}{2} E_{T \rightarrow 0} \left[ \frac{Z_T}{Z_{t+1}} e^{-w_{T \rightarrow t+1}} \ln \frac{2}{1 + \frac{Z_t+1}{Z_{t}} e^{-w_{t+1 \rightarrow t}}} \right],
\label{eq:JSD_FC}
\end{eqnarray}
where the first average is over the forward and the second average over the reverse process.

As in many of the above expressions, using this equation requires estimates of the partition function ratios.  Feng and Crooks suggested maximizing a log-likelihood, \cite{Feng2009}
\begin{eqnarray}
\mathcal L \left( \left\{ \frac{\hat{Z}_t}{\hat{Z}_{t'}} \right\} \right) & = & 
\sum_{t=0}^{T-1} 
\left[ \sum_{n=1}^{N_f} \frac{\hat{Z}_0}{\hat{Z}_t} e^{-w_{0 \rightarrow t}[X_{fn}]} 
		\ln \frac{1}{1+\frac{\hat{Z}_t}{\hat{Z}_{t+1}} e^{-w_{t \rightarrow t+1}[X_{fn}]}} \right] + \nonumber \\ & & 
		\sum_{t=0}^{T-1} 
\left[ \sum_{n=1}^{N_r} \frac{\hat{Z}_T}{\hat{Z}_{t+1}} e^{-w_{T \rightarrow t+1}[X_{rn}]} 
		\ln \frac{1}{1+\frac{\hat{Z}_{t+1}}{\hat{Z}_{t}} e^{-w_{t+1 \rightarrow t}[X_{rn}]}}
\right]
\label{eq:Likelihood_FC}
\end{eqnarray}
%\end{widetext}
We find, however, that this estimator does not perform well.  Starting with an estimate from Eq.~\ref{eq:estimator-for-normalization-constants}, we maximize $\mathcal L$ using a steepest descent method.  The norm of the gradient becomes nearly undetectable (less than $10^{-12}$) after only a few steps.  Unfortunately, the resulting estimate has an unreasonably large change between the first two and last two time points: the ratios $Z_0/Z_1$ and $Z_{T}/Z_{T-1}$ are very small.  This is because $Z_0$ or $Z_T$ are present in every sum of the expression and adjusting them has a disproportionate effect on the log-likelihood.  Because the log-likelihood is always negative (the exponential is always positive and the log fermi function always negative), values of $Z_0$ and $Z_T$ that are very small maximize the log-likelihood.  Convergence of Eq. \ref{eq:Likelihood_FC} would likely require an inordinate amount of data.

\begin{figure}[tb]
\includegraphics{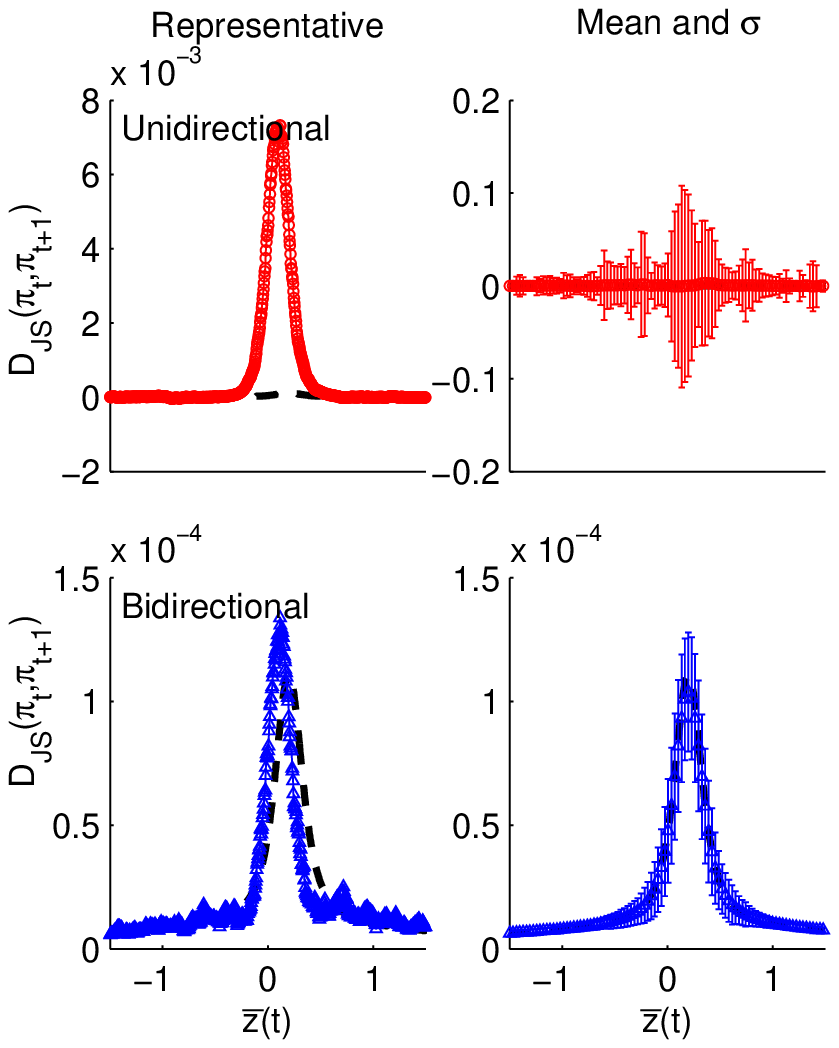}
\caption{{\bf Estimates of the Jensen-Shannon divergence.}
Representative estimates (left column) and the mean and standard deviation of 2500 estimates (error bars, right column) of the Jensen-Shannon divergence, $D_{JS}(\pi_t,\pi_{t+1})$, shown as a function of $\bar{z}(t)$ from $-1.5$ to $+1.5$.
Estimates were computed with the unidirectional, Eq.~\ref{eq:unidirectional-path-estimator} (top), or bidirectional, Eq.~\ref{eq:bidirectional-path-estimator}, estimator for the path-averages in Eq.~\ref{eq:JSD_FC} utilizing 125 pulling simulations in each direction.
For improved clarity, not all points are shown.
The value of $D_{JS}(\pi_t,\pi_{t+1})$ computed by numerical quadrature is shown as a thick dashed black line.
\label{fig:JSD}} 
\end{figure}

Because of the poor performance of Eq.~\ref{eq:Likelihood_FC}, we instead use the single time-slice estimator, Eq.~\ref{eq:estimator-for-normalization-constants}, in estimating Jensen-Shannon divergences using Eq.~\ref{eq:JSD_FC}.  Each path-ensemble average in Eq.~\ref{eq:JSD_FC} may be estimated with a unidirectional (Eq.~\ref{eq:unidirectional-path-estimator}) or bidirectional (Eq.~\ref{eq:bidirectional-path-estimator}) estimator.
The performance of the bidirectional estimator is vastly superior (See Fig.~\ref{fig:JSD}).
The unidirectional estimator strongly deviates from the reference value of the Jensen-Shannon divergence, especially around $-0.5<\bar{z}(t)<0.5$, with a mean and standard deviation that indicate extremely poor convergence.  
The cause of this poor convergence is likely similar to problems in unidirectional estimates from Jarzynski's equality: \cite{Jarzynski1997a,Jarzynski1997b}
rare events dominate estimates of exponential averages. \cite{Jarzynski2006}
Bidirectional estimates, on the other hand, attain much closer agreement with the true value.
While the Jensen-Shannon divergence and the variance are distinct quantities, they do bear considerable resemblance, and the performance of the bidirectional single time-slice estimator mirrors its performance in calculating the variance.

\begin{figure}[tb]
\includegraphics{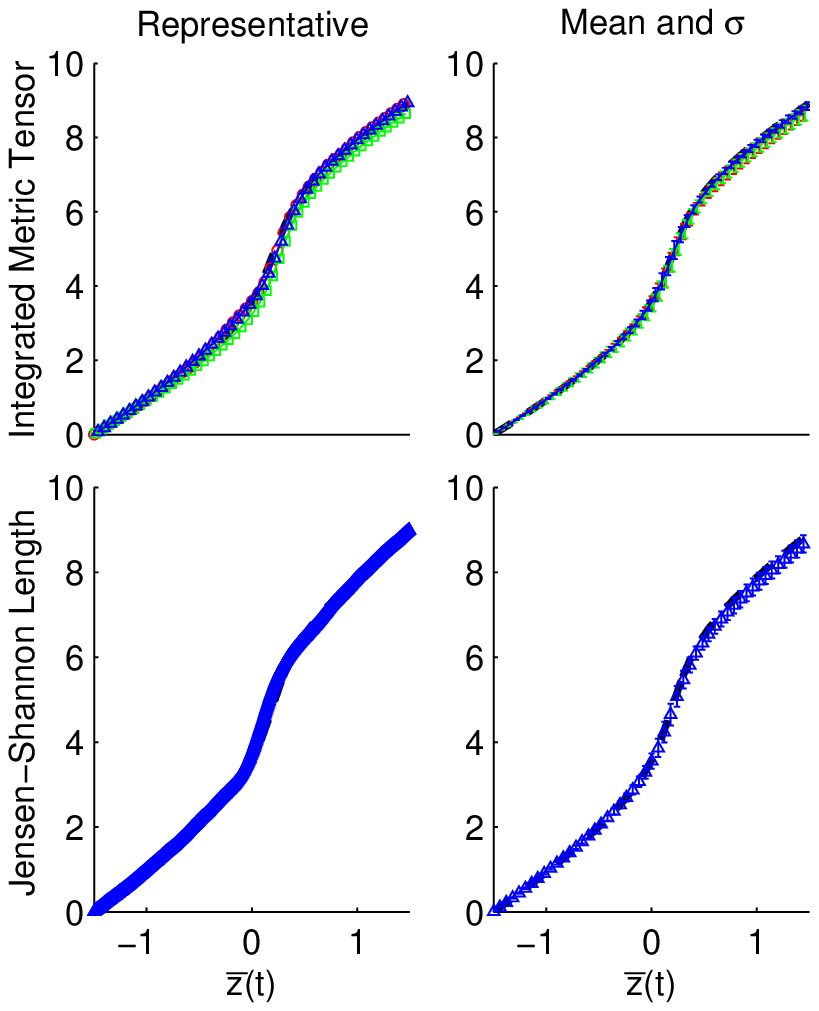}
\caption{{\bf Estimates of the thermodynamic length.}
Representative estimates (left column) and the mean and standard deviation of 2500 estimates (error bars, right column) of the thermodynamic length
between states with the harmonic bias centered around $\bar{z}(t) = -1.5$ and 750 values of $\bar{z}(t)$ (x-axis) up to $\bar{z}(t) = 1.5$, with each estimate based on 125 pulling simulations in both directions.
Estimates were either made using the trapezoidal rule with the metric tensor (top) or the Jensen-Shannon length (bottom).
Estimates of the metric tensor were computed with the single time-slice estimator, Eq.~\ref{eq:estimator-for-normalization-constants} (red circles or error bars), 
the MIS estimator with uniform weighting, Eq.~\ref{eq:MIS-estimator-for-normalization-constants-uniform} (green squares or error bars), 
and the MIS estimator with the balance heuristic, Eq.~\ref{eq:MIS-estimator-for-normalization-constants-balance} (blue triangles or error bars).
The Jensen-Shannon divergence was estimated with the bidirectional estimator, Eq.~\ref{eq:bidirectional-path-estimator}, to compute the path-averages in Eq.~\ref{eq:JSD_FC}, and length estimated with Eq.~\ref{eq:JSlength}.
For improved clarity, not all points are shown.
The thick black line shows the value of the thermodynamic length based on Gauss-Kronrod quadrature.
\label{fig:tL}} 
\end{figure}

\begin{figure}[tb]
\includegraphics{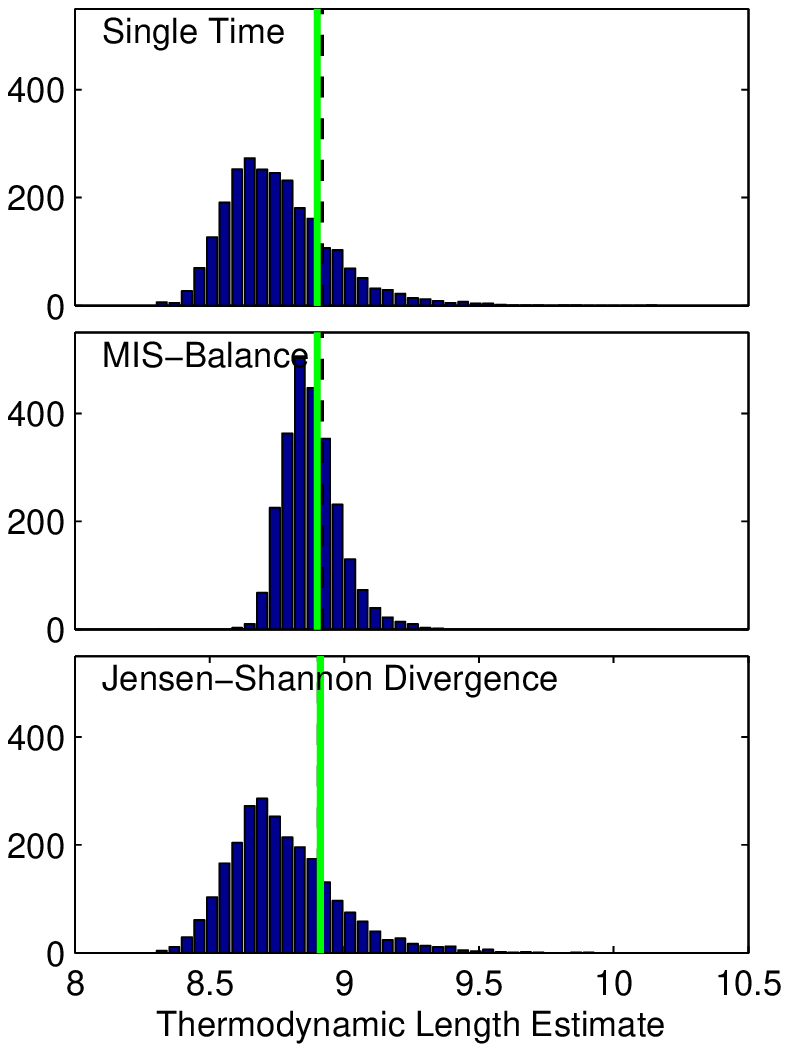}
\caption{{\bf Estimates of the total thermodynamic length.}
Histogram of estimates of the thermodynamic length from 2500 independent realizations of the pulling experiment using the single time-slice estimator, Eq.~\ref{eq:estimator-for-normalization-constants} (top) or the MIS estimator with the balance heuristic, Eq.~\ref{eq:MIS-estimator-for-normalization-constants-balance} (middle), to compute the metric tensor.  
The thermodynamic length $\mathcal{L}$ is then estimated using the trapezoidal rule.
For the histogram in the bottom panel, the Jensen-Shannon divergence is estimated with the bidirectional estimator, Eq.~\ref{eq:bidirectional-path-estimator}, to compute the path-averages in Eq.~\ref{eq:JSD_FC}, and the thermodynamic length estimated with Eq.~\ref{eq:JSlength}.
All simulations utilized 125 pulling simulations in each direction.
The thick black line shows the value of the thermodynamic length based on Gauss-Kronrod quadrature.
In the top two panels, the green line shows the thermodynamic length estimated based integrating the metric tensor at 750 points using Gauss-Kronrod quadrature and applying the trapezoidal rule to compute the length.
In the bottom panel, the green line is the Jensen-Shannon length.
\label{fig:tLhist}} 
\end{figure}

Hence, we have bidirectional estimators that, for our model system, perform reasonably well in estimating the metric tensor and the Jensen-Shannon divergence.  How do these estimators compare in computing the thermodynamic length?  In Fig.~\ref{fig:tL}, we compared some estimates of the thermodynamic length between states with the harmonic bias centered around $\bar{z}(t) = -1.5$ and 750 values of $\bar{z}(t)$ up to $\bar{z}(t) = 1.5$.
Of the methods using the metric tensor, the MIS estimator using the balance heuristic, as expected, performs the best.
The Jensen-Shannon length performs rather well but somewhat underestimates the thermodynamic length.
This is not a problem with discretization.
At this level of discretization, using the Jensen-Shannon length is very close to the thermodynamic length; indeed, it is superior than applying the trapezoidal rule to the metric tensor! (See Fig.~\ref{fig:tLhist}).

Trends in these estimates can be seen more clearly by a histogram of thermodynamic length estimates between states with the harmonic bias centered around $\bar{z}(t) = -1.5$ and $1.5$. (See Fig.~\ref{fig:tLhist}).
In this histogram, it is clear that estimates of the Jensen-Shannon length based on  Eq.~\ref{eq:JSD_FC}, as well as estimates of the thermodynamic length based on the single time-slice estimator of the metric tensor, do not perform as well as the MIS estimator for the metric tensor with the balance heuristic.
While the former two methods exhibit similar performance, the bias and variance of the latter estimator are substantially reduced.  
This improved performance reflects the aforementioned ability of the estimator to more accurately estimate the metric tensor at extreme values of $\bar{z}(t)$ (Fig.~\ref{fig:moments123}, middle).

%%%%%%%%%%%%%%%%%%%%%%%%%%%%%%%%%%%%%%%%%%%%%%%%%%%%%%%%%%%%%%%%%%%%%
\section{Discussion}
%%%%%%%%%%%%%%%%%%%%%%%%%%%%%%%%%%%%%%%%%%%%%%%%%%%%%%%%%%%%%%%%%%%%%

In this paper, we have described a stable method that generalizes previous estimators for potentials of mean force \cite{HummerSzabo2001,Oberhofer2009} to estimate arbitrary equilibrium expectations using multiple time slices from driven nonequilibrium processes.
While the estimator is not asymptotically efficient, we find that in our demonstrative simulations, using the balance heuristic of MIS leads to smooth and robust estimates of several properties with less bias and variance than other discussed estimators.
It is possible, however, that other choices of weights will lead to an estimator with even better properties.

With MIS, a good weighting function is proportional to the density from which the data are sampled.  
Thus, for sampling from multiple equilibrium distributions, the balance heuristic is provably good. \cite{Veach1995, Veach1997}
In driven nonequilibrium processes, however, samples from individual time slices are not drawn from the equilibrium density, but a nonequilibrium density which is likely closer to an equilibrium distribution earlier in the process; as mentioned previously, driven processes are known to exhibit \emph{lag}. \cite{Pearlman1989}
A weighting function which accounts for the lag could lead to an estimator with superior performance.
This is a possible future research direction.

Athen\`{e}s and Marinica \cite{Athenes2010} have proposed a different estimator that pools data from multiple time slices of a driven nonequilibrium processes to estimate equilibrium expectations.  
Their strategy entails using a Bayesian posterior (with an equilibrium prior) for the probability of observing a position during an entire trajectory.
As their method was developed in the context of biased path sampling, it is not directly relevant to the situations described in this paper.
A future comparison of the two strategies will likely require developing their strategy into a new estimator.

We conclude by noting that the MIS strategy may not only be used for combining data from multiple time slices from driven nonequiliibrium processes, but for pooling data from both equilibrium and nonequilibrium data.
We expect that this feature will be useful in the context of enhanced equilibrium sampling methods that use driven nonequilibrium processes to generate trial moves for Monte Carlo simulations. \cite{Opps2001,Ballard2009}

%%%%%%%%%%%%%%%%%%%%%%%%%%%%%%%%%%%%%%%%%%%%%%%%%%%%%%%%%%%%%%%%%%%%%
\section{Acknowledgments}
%%%%%%%%%%%%%%%%%%%%%%%%%%%%%%%%%%%%%%%%%%%%%%%%%%%%%%%%%%%%%%%%%%%%%

The authors thank David Sivak (LBNL), Gabriel Stoltz (CERMICS, Ecole des Ponts ParisTech), and Attila Szabo (NIH) for insightful discussion and helpful feedback.
DDLM acknowledges support from a Director's Postdoctoral Fellowship at Argonne and JDC from a QB3-Berkeley Distinguished Postdoctoral Fellowship.

%\bibliography{neqMIS}
%Merlin.mbs v4.21 2009-07-09.
%

\end{document}